# st-DTPM: Spatial-Temporal Guided Diffusion Transformer Probabilistic Model for Delayed Scan PET Image Prediction

Ran Hong, Yuxia Huang, Lei Liu, Zhonghui Wu, Bingxuan Li, Xuemei Wang, Qiegen Liu, *Senior Member, IEEE*

*Abstract*—PET imaging is widely employed for observing biological metabolic activities within the human body. However, numerous benign conditions can cause increased uptake of radiopharmaceuticals, confounding differentiation from malignant tumors. Several studies have indicated that dual-time PET imaging holds promise in distinguishing between malignant and benign tumor processes. Nevertheless, the hour-long distribution period of radiopharmaceuticals post-injection complicates the determination of optimal timing for the second scan, presenting challenges in both practical applications and research. Notably, we have identified that delay time PET imaging can be framed as an image-to-image conversion problem. Motivated by this insight, we propose a novel spatial-temporal guided diffusion transformer probabilistic model (st-DTPM) to solve dual-time PET imaging prediction problem. Specifically, this architecture leverages the U-net framework that integrates patch-wise features of CNN and pixel-wise relevance of Transformer to obtain local and global information. And then employs a conditional DDPM model for image synthesis. Furthermore, on spatial condition, we concatenate early scan PET images and noisy PET images on every denoising step to guide the spatial distribution of denoising sampling. On temporal condition, we convert diffusion time steps and delay time to a universal time vector, then embed it to each layer of model architecture to further improve the accuracy of predictions. Experimental results demonstrated the superiority of our method over alternative approaches in preserving image quality and structural information, thereby affirming its efficacy in predictive task.

*Index Terms*—Dual-time PET imaging, DDPM, transformer, spatial-temporal guided.

This work was supported in part by Chinese National Natural Science Foundation (82271267, 62201193), and the Natural Science Foundation of Guangdong Province (2022A1515011670). (Corresponding authors: X. Wang and Q. Liu.) (R. Hong, Y. Huang and L. Liu are co-first authors.)
This work did not involve human subjects or animals in its research.

R. Hong, Y. Huang, Z. Wu and Q. Liu are with School of Information Engineering, Nanchang University, Nanchang 330031, China. ({ranhong, 416100220082, wuzhonghui}@email.ncu.edu.cn, liuqiegen@ncu.edu.cn)
L. Liu is with Department of Nuclear Medicine, Affiliated Hospital of Inner Mongolia Medical University, Huhhot North Street, Inner Mongolia, 010050, China. (liulei_1124@126.com)
B. Li is with Institute of Artificial Intelligence, Hefei Comprehensive National Science Center, Hefei 230088, China. (libingxuan@iai.ustc.edu.cn)
X. Wang is with Department of Nuclear Medicine, Affiliated Hospital of Inner Mongolia Medical University, Huhhot North Street, Inner Mongolia, 010050, China and Department of Nuclear Medicine, The First Affiliated Hospital of USTC, University of Science and Technology of China, Hefei, Anhui 230001, China. (wangxuemei201010@163.com)

## I. INTRODUCTION

18F-fludeoxyglucose positron emission tomography （18F-FDG PET/CT) imaging is the most widely used and valuable molecular imaging technology in cancer clinic. The sensitivity of tumor diagnosis can reach 96% (83% to 100%), the specificity is 79% (52% to 100%), and the accuracy is 91% (86% to 100%) [1]. However, with the increasing clinical application of 18F-FDG tumor imaging, more and more false positive lesions have been found. Including inflammatory lesions (such as bacterial pneumonia, pulmonary abscess, aspergillosis) and granulomatous lesions (such as tuberculosis, sarcoidosis, histoplasmosis, Wegener granuloma, silicosis, etc.). These diseases make 18F-FDG PET/CT imaging difficult in the differential diagnosis of benign and malignant tumor lesions [2]. The main reason is that 18F-FDG is an analogue of glucose, which can be taken up by tumor cells or mononuclear macrophages, resulting in false positive results. Studies [3-5] have found that the peak time of 18F-FDG uptake by tumor cells and inflammatory cells is different. The peak time of 18F-FDG uptake by tumor cells is 3-4 hours after 18F-FDG injection, and the peak of uptake by inflammatory lesions is about 1 hour. Therefore, how to improve the efficiency of 18F-FDG PET/CT in the identification of benign and malignant tumors through the improvement of image acquisition methodology is a problem worth studying at present. Among them, delayed scan PET imaging (or dual-time PET imaging) technology is an effective method to improve the identification of benign and malignant tumors. The principle of delayed PET images acquisition is to take advantage of the different uptake or excretion levels of imaging agents in various tissues and cells at different time periods. A conventional image is collected about 1 hour after 18F-FDG injection, and then a delayed image is collected 2 hours after developer injection. Biological differences between tumors, inflammation, and normal tissues with high physiological uptake are identified by comparing changes in levels of imaging agent uptake or excretion over time periods. Currently, in many clinical studies of breast cancer, lung cancer, rectal and prostate cancer, etc., it has been found that standard uptake value (SUV) can significantly improve the accuracy of differentiating benign and malignant tumors through delayed PET images acquisition and quantitative analysis of lesions [6-10].

However, in the field of PET delayed imaging research, there are some key challenges. 1) Increase the localized radiation dose for patients; 2) Prolonged patient stays in cory pose an increased clinical risk, particularly for critically ill patients, while also adding to the psychological burden. 3) Occupying a certain amount of time reduces machine turnover and total inspector count. 4) Extend staff working hours. 5) Raise staff radiation exposure levels. 6) Expand image storage capacity. These significant obstacles have led to a lack of research on PET delayed imaging prediction. At present, some studies have made progress in predicting the most appropriate time for delayed scanning, but they have not yet addressed the underlying issues [11, 12].

With the availability of extensive training datasets and substantial computational resources, the excellent performance of deep learning in the medical image domain has received increasing attention in recent years [13]-[15]. Generative adversarial network (GAN) is driven by adversarial learning of a pair of generator and discriminator. It presents many impressive performances across a series of image-to-image tasks such as image denoising [16], style transform [17], inpainting [18], and super-resolution [19]. Those GAN based image-to-image tasks are introduced by pix2pix. Image-to-image translation is the process of utilizing deep learning architectures in the field of computer vision to convert images from one domain to another by capturing shared correlations and feature representations between the two image types. For example, it can convert grayscale images to color images [20], translate satellite images into maps [21], or transform oil paintings into realistic photographs [22]. Additionally, diverse applications exist for this technology, including medical image reconstruction, precise image segmentation, advanced image enhancement, and intricate cross-domain image transformation [23, 24]. Surprisingly, we find that PET delayed imaging can also be formulated as an image-to-image conversion problem, where a model needs to be found to convert the first scan PET image into a delayed scan PET image. However, directly estimating a delayed PET image from the first scan also poses technical challenges, as some benign lesions such as inflammation and nodular granuloma can also have tracer uptake, complicating accurate diagnosis.

In the field of delayed imaging technology, few methods for generating delayed PET images have been reported. As far as we know, the generation of delayed PET images using deep learning methods has not been validated. In fact, due to the broader dynamic range of organ uptake and varying noise levels in scan results, the data distribution of PET images is more intricate than that of natural images. Notably, the work [25, 26] built a GAN model that includes a U-net generator and patch discriminator to obtain certain performance in the task which similar to ours. However, both models suffer from weak generator capabilities and shortcomings in adversarial learning, which hinder their ability to accurately predict PET images with long delays and achieve good performance on complex human structures.

The emergence of diffusion model [27]-[29] has brought generative model to a new stage and achieved remarkable results in some important fields. Among these, the denoising diffusion probability model (DDPM) [30] stands out as a representative model capable of enhancing information extraction from low-resolution PET images. To address the challenge of determining the timing for a delayed scan, we propose a novel diffusion model called spatial-temporal guided diffusion transformer probabilistic model for dual-time PET imaging (st-DTPM), aimed at alleviating clinical workload and enhancing diagnostic efficiency. The model is based on the U-net framework and combines the characteristics of CNN and pixel-wise transformer. It also adopts DDPM model under specific conditions to perform image diffusion. This study seeks to generate a delayed scan PET image by utilizing the first scan PET image via the st-DTPM model. Subsequently, we will assess the predictive accuracy of the st-DTPM model by comparing the generated results with the actual delayed PET images. Therefore, it can overcome the shortcomings of delayed scan PET imaging and make delayed PET images prediction more widely used in clinical practice.

The contributions in this paper are summarized as follows:

**Delayed PET imaging prediction based on spatial-temporal.** Building upon the DDPM model, we concatenate noisy delayed scan PET images with original first scan PET images across channels to establish an image-to-image mapping. The similarity between two samples under different delays is taken as an additional constraint of the model. Furthermore, we embed time step and delay time as vectors into the model framework to further enhance predictive performance.

**Diffusion transformer probabilistic model.** The architecture combines the advantages of CNN and transformer. We propose a conditional diffusion model based on U-net framework by mixing CNN and pixel-wise transformer, and embed different combinations of diffusion time step and delay time in both residual block and transformer block. Our approach is the first to apply deep learning to PET delayed imaging technology.

The remainder framework of this paper is outlined as follows. Section II briefly overviews some related works. Section III contains the key idea of the ST-DTPM approaches. The experimental settings shown in Section IV. Section V exhibits some experimental results. Section VI conducts a concise discussion and Section VII draws a conclusion for this work.

## II. RELATED WORK

### A. Denoising Diffusion Probabilistic Model

Diffusion model [30-32] is a Markov chain that gradually infuses noise into pure images and then progressively removes noise, transforming Gaussian noise into generate images. The process of injecting noise into raw images can be represented as the forward diffusion process:

$$q(x_t | x_{t-1}) = \mathcal{N}(x_t; \sqrt{\alpha_t} x_{t-1}, \beta_t \mathbf{I}) \quad (1)$$

here $\beta_t$ and $\alpha_t$ represent the schedule of injecting noise into $x_0$, the variance $\beta_t \in (0,1)$ increase with $t \sim uniform(\{1,\cdots,T\})$. The definition of $\alpha_t = 1-\beta_t$ and $\bar{\alpha}_t = \prod_{i=1}^{t}\alpha_i$, so $\alpha_t$ will decrease with $t$, and $\bar{\alpha}_t$ will decrease to $0$ as $t$ becomes large. Eq. (1) represents one step in the diffusion process of Markov chain.

$$q(x_t | x_0) = \mathcal{N}(x_t; \sqrt{\bar{\alpha}_t} x_0, (1-\bar{\alpha}_t)\mathbf{I}) \quad (2)$$

Since the parameters of the diffusion process are fixed, Eq. (1) can be extended to $t$ steps, then we derive Eq. (2), which means get a noise image from a pure image. The $x_t$ obeys a Gaussian distribution with mean $\sqrt{\bar{\alpha}_t}$ and variance $1-\bar{\alpha}_t$ for an arbitrary time step, it follows a normal distribution when $t$ is large enough. The backward reconstruction process involves removing noise from a normal distribution using a Markov chain, which reverses the diffusion process by Bayes' theorem. Then the optimal mean $\mu_\theta(x_t, x_0)$ will be:

$$\mu_\theta(x_t, x_0) = \frac{1}{\sqrt{\alpha_t}}(x_t - \frac{(1-\alpha_t)}{\sqrt{1-\bar{\alpha}_t}}\varepsilon_\theta) \qquad (3)$$

The noise $\varepsilon_\theta \sim \mathcal{N}(0, \mathbf{I})$ in each forward diffusion process needs to be predicted using a neural network. DDPM has constructed a deep learning model with embedded time step and U-net framework. This model takes the noisy image and time step as input to remove noise injected during the forward process. It can be seen as a series of denoising autoencoders, so the optimization target can be expressed as:

$$\min E_{t,x_0,\varepsilon}[\|\varepsilon - \varepsilon_\theta(\sqrt{\bar{\alpha}_t}x_0 + \sqrt{1-\bar{\alpha}_t}\varepsilon, t)\|^2] \qquad (4)$$

In prior work on diffusion model, a prevalent strategy involved employing CNN blocks to extract features and utilizing a U-net framework to enhance resolution for generating multi-scale feature maps. Recently, CNN-based models have faced challenges from transformer architectures. Typical diffusion transformer (DiT) [33] is a plain model which build by a series of transformer blocks. It discards the inductive bias of CNN but obtains the long-range relativity among every patches. It has achieved great success in image generation based on diffusion models.

### B. Vision Transformer

The transformer architecture was initially developed for text generation tasks, leveraging its core multi-head self-attention mechanism to effectively capture correlations between tokens [34]. Building upon this foundation, the Vision Transformer (ViT) [35] extends the transformer paradigm to the realm of computer vision. ViT adopts a patch-based approach, breaking down images into smaller patches and treating each patch as a token. Furthermore, position embeddings are added to each token prior to their input into the multi-layer transformer encoder blocks. This strategy enables ViT to effectively process visual data by leveraging the hierarchical representation learning capabilities inherent in the transformer architecture. ViT presents a compelling approach to capturing long-range correlations within images through the utilization of a pure transformer architecture. However, this method sacrifices the multi-scale information and robust local feature extraction capabilities inherent in CNN. Moreover, the choice of patch size in ViT is pivotal, as smaller patch sizes ($2 \times 2$) exponentially escalate computational requirements, while larger patch sizes ($32 \times 32$) compromise the ability of model to focus attention within individual patches. Addressing these limitations, the Swin-Transformer [36] merges design principles from both ViT and CNN architectures to offer a comprehensive solution. By employing patch merging techniques, Swin-Transformer enables down sampling of images, facilitating the acquisition of multi-scale information crucial for comprehensive feature extraction. Additionally, the integration of shifted window mechanisms allows the model to dynamically adjust the focus of attention within each patch, thereby enhancing the model's capacity for local attentional processing. Through these innovations, Swin-Transformer represents a significant advancement in image processing, combining the strengths of both ViT and CNN architectures while mitigating their respective weaknesses. However, the transformer architecture exhibits a high demand for data and demonstrates superior performance over CNN in the presence of extensive training datasets. The inductive bias of CNN confers an advantage when confronted with limited datasets.

## III. METHOD

### A. Motivation

The PET delayed imaging technique is more flexible than traditional PET imaging methods. As depicted in Fig. 1, in order to obtain a more accurate understanding of the distribution of radioactive drugs in the body, the subject needs to undergo multiple scans at different time points after injection. Clinical doctors can evaluate the distribution changes of the radioactive drugs in the body by comparing the images at the two different time points, which can help determine the activity of lesions or tumors.

Several studies have demonstrated that there is a certain correlation between the PET images of the first scan and the PET images of the delayed scan [37, 38]. Specifically, both scans involve PET imaging of tissue from the identical diseased site on the same subject that shares related information. Over time, tissue uptake of inflammation and other false positive lesions will gradually decrease, so the information contained will gradually decrease. The first scan provides information about molecular metabolism at the initial time point, while the delayed scan provides information about the time point of subsequent molecular changes. These temporal differences facilitate assessment of metabolic dynamics, such as tumor progression or treatment response.

However, since the radioactive drug needs at least one hour to distribute throughout the body after injection, the second scan time point is often difficult to determine, which is a heavy task in practical applications and research. Therefore, we are contemplating if there are alternative methods to expedite the acquisition of the delayed scan PET images. Due to the broader dynamic range across various organs and the varying noise levels in different scan results, the data distribution of PET images is more intricate and possesses lower resolution compared to natural images. However, it is worth noting that PET delay imaging can also be formulated as an image-to-image conversion problem, where a model needs to be found to convert a PET image from the first scan to delay scan. Therefore, we can obtain more characteristic information by mining their prior information [39]. Diffusion model leverages the powerful generation capabilities to demonstrate their unique value in various fields. DDPM stands out as a representative model capable of enhancing information extraction from low-resolution PET images. Inspired by such methods, we try to generate delayed scan PET images using diffusion model methods. Therefore, we propose a novel spatial-temporal diffusion transformer probabilistic model (st-DTPM) to solve delayed scan PET imaging prediction problem.

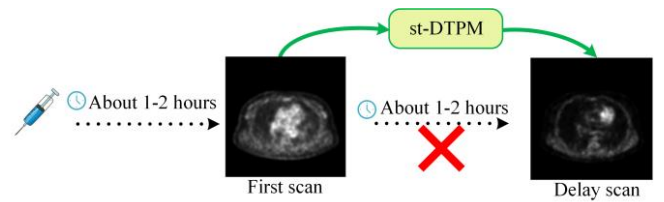

**Fig. 1.** A visualization of delayed scan PET imaging. A deep learning model is proposed to predict delayed PET scans from early PET scans, thereby reducing the waiting time for patients undergoing delayed imaging.

### B. Spatial-Temporal Conditional PET Imaging Prediction

In this section, we introduce a diffusion model st-DTPM that follows the U-net framework. This model combines the advantages of CNNs and ViTs to excavate deeper feature information. Additionally, we propose a delay time embedding method based on dual-time PET imaging prediction, which utilizes the idea of position encoding in [35] to convert the time value into a general time vector. Drawing inspiration from [40], we propose a hybrid structure shown in Fig. 2. At each stage of the U-net down sampling, there are two blocks: a multi-CNN block and a pixel-wise transformer block, which will be discussed in detail in Section III.C. Furthermore, to further enhance the prediction performance, we introduce the delay time embedding method, which is explained in following segment.

*Diffusion Model with Spatial Condition.* The overall concept of forward-backward process and the deep denoising model is illustrated in Fig. 2. The training and generation of this approach are followed by DDPM, which gradually injects Gaussian noise at discrete time steps and attempts to mitigate noise using a deep denoising model to generate images from standard Gaussian noise. Since dual-time PET imaging prediction leverages early PET scan images to predict delayed PET scan images, the fundamental principle is related closely to the image-to-image methodology. Therefore, we need to achieve mapping between paired images. For this purpose, during training process, we adopt a simple approach follows [41, 42], using pairs of early scan PET images $x_e$ and noisy delayed scanned images $x_t$ as model inputs. This method effectively facilitates the establishment of image-to-image mappings. We solve this task by adopting DDPM model for conditional image generation. The target of optimization in Eq. (4) can be reformulated as:

$$\min E_{t,x_0,\varepsilon}[\|\varepsilon - \varepsilon_\theta(Concat(\sqrt{\bar{\alpha}_t}x_0 + \sqrt{1-\bar{\alpha}_t}\varepsilon, x_e), t)\|^2] \quad (5)$$

where *Concat* is an operator that denotes the concatenation of pair images in channel dimension. In this way, a one-to-one image mapping can be achieved.

$$c = Concat(x_e, x_t) \quad (6)$$

where $c$ denotes spatial condition that embeds to the model.

In the context where of early scan PET image is used to as a condition for the denoising process, the backward processing proceeds as follows:

$$q(x_{t-1} | x_t, c) = \mathcal{N}(x_{t-1}; \mu_\theta(x_t, c), \sigma_t \mathbf{I}) \quad (7)$$

$$\mu_\theta(x_t, c) = \frac{1}{\sqrt{\alpha_t}}(x_t - \frac{\beta_t}{\sqrt{1-\bar{\alpha}_t}}\varepsilon_\theta(x_t, c)) \quad (8)$$

It can be seen from Eqs. (7) and (8) that conditional control is solely related to deep denoising models. Therefore, we use a method similar to [43, 44] which concatenate noisy images $x_t$ and condition image $x_e$ on channels before every reconstruction step. To learn conditional diffusion models, such as class-conditional [45] or text-to-image [46] models, the conditional information is additionally incorporated into the noise prediction objective. Ultimately, the optimization target can be reformulated as:

$$\min E_{t,t_d,c_t,\varepsilon}[\|\varepsilon - \varepsilon_\theta(\sqrt{\bar{\alpha}_t}c + \sqrt{1-\bar{\alpha}_t}\varepsilon, t, t_d)\|^2] \quad (9)$$

where $t_d = st(x_0) - st(x_e)$ is the delay time interval that the difference between first scan time and delayed time scan.

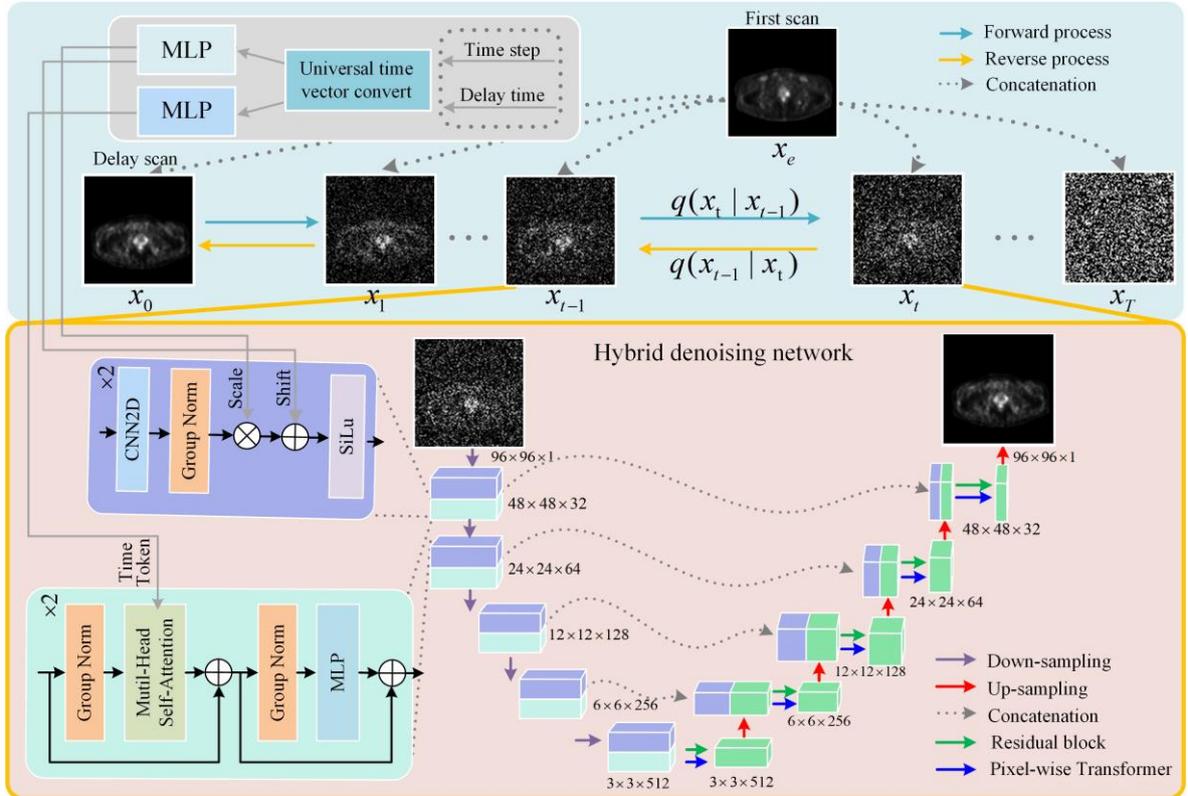

**Fig. 2.** The overall concept of the proposed st-DTPM model. Top left is temporal guided which convert time step and delay time to vectors, then embed them into every block of deep denoising model; Top right is spatial guided which concatenation first scan image and delay scan image on channel; Middle part is diffusion and reconstruction processes following DDPM framework; Lower part of the module illustrates the hybrid deep denoising model.

*Diffusion Model with Temporal Condition.* The diffusion model can be conceptualized as a cascade of denoising autoencoders. Each denoising autoencoder receives a noise image $x_t$ as input and outputs its corresponding denoised image $x_{t-1}$. The level of noise in the input image decreases incrementally as it passes through the autoencoder. Instead of employing a series of denoising autoencoders to handle multi-level noise, the diffusion model utilizes a universal denoising autoencoder. Consequently, it is essential to ensure that the model accurately discerns the noise level present in the current input image during the denoising process within the Markov chain, enabling precise completion of the denoising task. Our diffusion model framework is similar to DDPM [40, 47], which belongs to the discrete time diffusion model. The time step is set to 300, thus we denote time step $t \in \{0, 300\}$.

Delay time of dual-time PET imaging greatly affects the gap of radioactive quantification between first scan and delay scan. Based on this, we believe that delay time is a crucial factor affecting PET images. Therefore, we incorporate delay time as a temporal constraint within our model to further enhance prediction accuracy. Specifically, we embed the delay time using the same approach as the time step embedding in the diffusion model. Considering that the delay time interval $t_d$ is a continuous variable and in order to transform it into a universal time vector, we need to convert it into a discrete time variable. Therefore, we ignore seconds and only use minutes to represent the delay time interval.

In order to unify the scale of two types of time information, we need to convert $t$ and $t_d$ into universal time vector $T \in \mathbb{R}^{N \times 1 \times 1}$ and $T_d \in \mathbb{R}^{N \times 1 \times 1}$, respectively. Then embed them into the multi-CNN block and pixel-wise transformer block in two different ways. Universal time vector converter is depicted in Fig. 3, it consists of sinusoidal position encoding follows [34] and bottleneck MLP. The sinusoidal position encoding uses the superposition of sine and cosine waves to transform discrete position or time into a vector $T_S \in \mathbb{R}^N$. It is illustrated as follows:

$$PE_{(pos, 2i)} = \sin(pos / 10000^{2i/N}) \quad (10)$$

$$PE_{(pos, 2i+1)} = \cos(pos / 10000^{2i/N}) \quad (11)$$

where $N$ represents the length of the universal time vector, and $pos$ represents the discrete time step of our input. The bottleneck MLP is built by dimension expansion linear layer and dimension compression linear layer, where the former expand dimension of $T_s$ to $T_e \in \mathbb{R}^{4N}$, and the latter compression the dimension of $SiLU(T_e)$ to $T_u \in \mathbb{R}^N$.

There are two universal time vectors: the diffusion model time step vector $T \in \mathbb{R}^{N \times 1 \times 1}$ and the delay time vector $T_d \in \mathbb{R}^{N \times 1 \times 1}$, which need to be embedded into the model. To explore the interaction between two vectors, we have considered four combinations before embedding them into the model: 1) Each block (EC): alternately embed $T$ and $T_d$ into each block; 2) Linear adding (LA), adding them together and performing a linear projection i.e., $linear(T + T_d)$. 3) Linear concatenation (LC): concatenating them and performing a linear projection i.e., $linear(\mathrm{C}oncat(T, T_d))$; 4) Adding (AD): directly adding them i.e., $T + T_d$; Among them, the first approach combines effectiveness and efficiency. We will conduct a comprehensive analysis of the distinctions among these four embedding methodologies in ablation study.

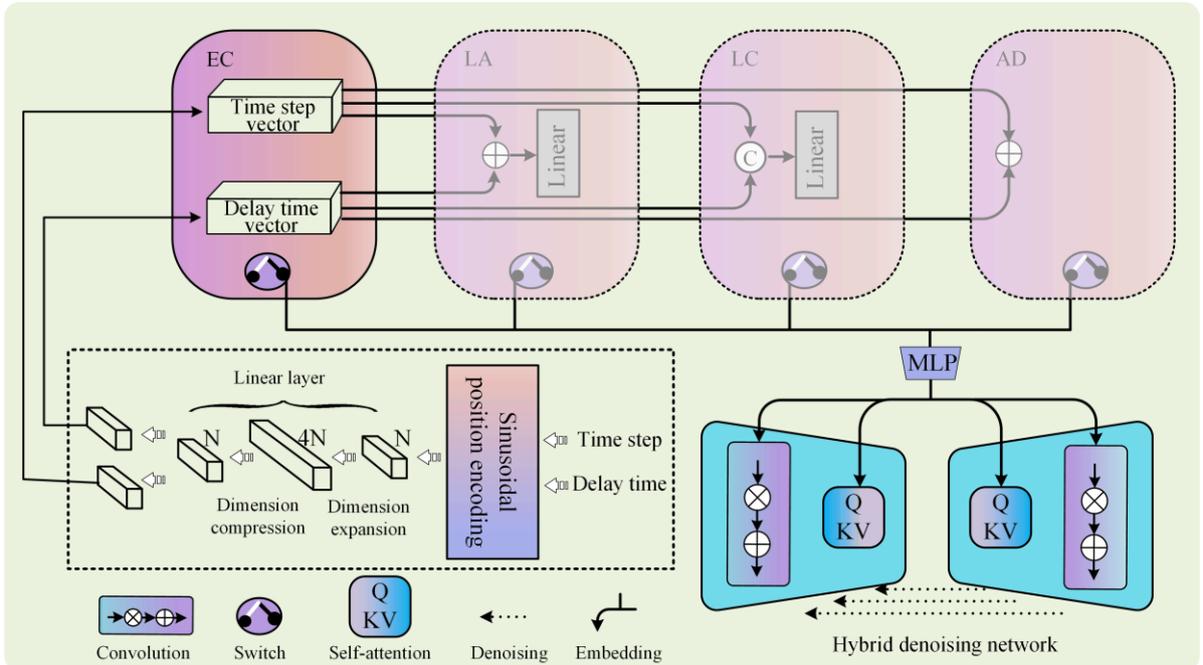

**Fig. 3.** Four various universal time vector embedding methodologies. We first transform the time step and delay time into universal time vector. Then, the switch operation is used to select the most appropriate time vector embedding method (EC). Finally, after the MLP process, the universal time vector will embed time condition into residual block and transformer block.

*C. Denoising Network Combining CNN and Transformer*

As illustrated in Fig. 2, the denoising module of the st-DTPM model consists of multiple stages. Each stage consists of the following two parts: (1) a cascade of residual block which obtains multi-scale visual feature and makes model possess inductive bias. (2) a pixel-wise transformer block which gives feature map long range relative among every pixel. PET images are characterized by their inherently blurred structure and the correlations between different regions. In one case, if a model composed solely of convolutional layers fails to capture the mutual influence of SUV values across different anatomical regions and cancerous tissues. In another case, when the model is built only with transformer, due to the lack of inductive bias and in the case of small data volume, relying solely on pixel correlations makes it difficult for the model to reconstruct the inherent human body shape in PET images. Therefore, we consider combining these two models to take full advantage of their strengths.

Our proposed model integrates multi-scale information through multi-level sampling. It leverages convolutional layers to extract local structural details at each feature scale and employs transformer to effectively capture regional correlations. This hybrid approach enhances the model's capability to denoise PET images and reconstructs high-quality PET images, thereby improving diagnostic accuracy and clinical relevance.

***Convolution Layer.*** The module is followed by [48] which consists of convolution layer ($K=3\times3$, $S=1$, $P=1$), group normalization ($Group=4$) and *SiLU* activation function. The operation between group normalization and the activation function serves as the time embedding mechanism, aiming to embed the time step from the denoising process of the diffusion model and the time interval condition from dual-time PET imaging into each block of the model. Specifically, we denote the input feature map as $M \in \mathbb{R}^{C\times H\times W}$, where $H$, $W$ and $C$ represent the height, width, and channels of the feature map, respectively. Due to the hyperparameter setting of convolution layers, the shape of feature map is still keeping $M \in \mathbb{R}^{C\times H\times W}$. Before the universal time vector is embedded in the convolution layer, we perform a linear projection that is denoted as $MLP(T_u \in \mathbb{R}^{2C\times 1\times 1})$, then we chunk it into scale vector and shift vector, where $scale, shift \in \mathbb{R}^{C\times 1\times 1}$. Eq. (12) will embed time condition into residual block.

$$M_T = (scale+1)\times M + shift \quad (12)$$

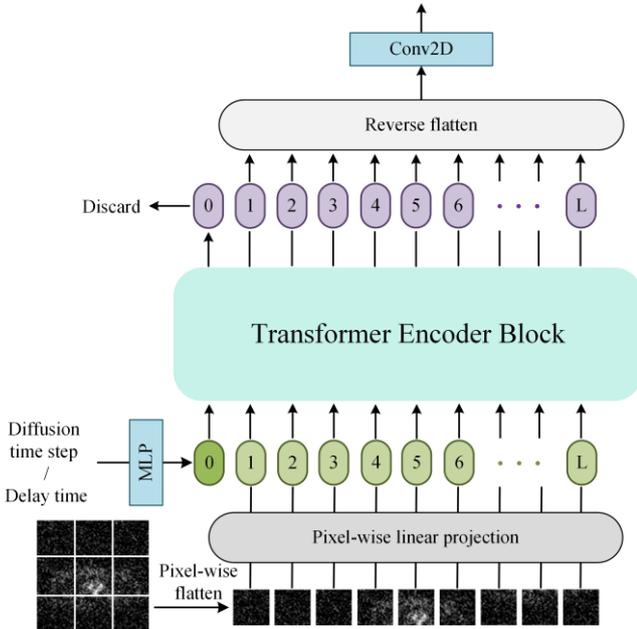

**Fig. 4.** Overview of the pixel-wise transformer layer. The time variable serves as a token within the transformer, with the token comprising noisy image pixels being input into the transformer encoder block. After assigning attention weights to each pixel in the image, the delay time token is removed and the pixel matrix is reconstructed.

***Pixel-wise Transformer Layer.*** The module follows ViT [35]. First, the given feature map $x_h \in \mathbb{R}^{C\times H\times W}$ is passed through a flatten layer to obtain a pixel vector. Then, this vector undergoes a pixel-wise linear projection to generate a sequence of tokens $x_s \in \mathbb{R}^{L\times D}$, where $L=H\times W$, $D$ is hidden dimension of transformer block. Due to the small size of the original PET images, we adopt a pixel-based transformer, which can obtain correlations between each pixel without adding too much computational complexity. On this basis, we have added time condition to further improve its predictive accuracy. Specifically, the query $\mathbf{Q}$, key $\mathbf{K}$ and value $\mathbf{V}$ of multi-head self-attention are redefined as Eqs. (13), (14) and (15).

$$\mathbf{Q}_s = Concat(W_{qs}\times x_s, W_t \times T_u) \quad (13)$$
$$\mathbf{K}_s = Concat(W_{ks}\times x_s, W_t \times T_u) \quad (14)$$
$$\mathbf{V}_s = Concat(W_{vs}\times x_s, W_t \times T_u) \quad (15)$$

where $T_u$ is a universal time vector processed by a linear layer, whose length equals the length of the tokens. $W_{qs}$, $W_{ks}$, $W_{vs}$ denote pixel-wise linear projection weights for $\mathbf{Q}$, $\mathbf{K}$ and $\mathbf{V}$. $W_t$ represents temporal linear projection weights for universal time vector. The self-attention is computed as follows:

$$Attention = Softmax\left(\frac{\mathbf{Q}_s \times \mathbf{K}_s^T}{\sqrt{D}}\right)\times \mathbf{V}_s \quad (16)$$

By injecting the time token into the attention calculation, each pixel in the feature map becomes correlated with the temporal condition. This correlation enables the model to adjust the denoised SUV with respect to time throughout the denoising process. After transformer block, the time token is discarded, the time token is discarded, reverting the feature map to its original arrangement. The temporal condition transformer is defined as:

$$\hat{x}_s = MSA(W_S \times Concat(GN(x_s)+W_t\times x_t))+x_s \quad (17)$$
$$x_s = MLP(GN(UnConcat(\hat{x}_s, \hat{x}_t)))+\hat{x}_s \quad (18)$$

where *MSA* denotes multi-head self-attention which is described as Eq. (17), *MLP* and *GN* denote multi-linear projection and group norm, respectively.

Algorithm 1 and 2 provide a detailed explanation of the delayed PET image prediction process.

---

**Algorithm 1:** Training

1: **repeat**
2: $(x_0, x_e) \sim p(x, x_e)$
3: $t \sim uniform(\{1,\cdots,T\})$
4: $t_d = st(x_0)-st(x_e)$ (Delay time interval)
5: $\varepsilon \sim \mathcal{N}(0,\mathbf{I})$
6: **take a gradient descent step on**
$$\nabla_\theta \left\|\varepsilon-\varepsilon_\theta(Concat(\sqrt{\bar{\alpha}_t}x_0+\sqrt{1-\bar{\alpha}_t}\varepsilon, x_e), t, t_d)\right\|^2$$
7: **until converged**

**Algorithm 2:** Prediction

1: $x_T \sim \mathcal{N}(0,\mathbf{I})$
2: $(x_e) \sim p(x_e)$
3: **for** $t=T,\cdots,1$ **do**
4: $z \sim \mathcal{N}(0,\mathbf{I})$ **if** $t>1$, **else** $z=0$
5: **channel concatenation**:
$$c = Concat(x_e, x_t)$$

| | | |
|---|---|---|
| 6: | $x_{t-1} = \frac{1}{\sqrt{\alpha_t}}(x_t - \frac{1-\alpha_t}{\sqrt{1-\bar{\alpha}_t}}\varepsilon_\theta(c,t,t_d)) + \sigma_t z$ | |
| 7: | **end for** | |
| 8: | **return** $x_0$ | |

## IV. EXPERIMENT

### A. Datasets Details

The dataset used in this study is clinical data from 232 patients who undergo PET/CT routine examination and delayed 18F-FDG PET/CT scan at the affiliated hospital of Inner Mongolia medical university from January 2018 to December 2023, there are 122 cases of prostate cancer and 110 cases of lung cancer, and all patients' sensitive information is erased. All 233 patients have obvious lesions and are diagnosed through biopsy, surgical resection, or follow-up for more than 12 months.

During the PET/CT data acquisition phase, the patient's fasting blood glucose level is controlled below 11.1 $mmol/L$ before the scan. The patient needs to lie down and rest for 40-60 minutes, after inject 18F-FDG intravenously at a dose of 0.15 $mci/kg$. Then, empty the bladder, drink 500 $ml$ of milk, and perform PET/CT imaging. During imaging, the patient is placed in a supine position with both arms raised to the top of the head, and the scanning range is from the skull to the upper middle femur. During CT scanning, CARE Dose4D is used as the current reference, with a reference current of 350 $mA$, a voltage of 120 $kV$, and a reconstructed layer thickness of 5.0 $mm$. Then, a PET scan is performed with a scanning speed of 1.7 $mm$/s and FWHM 5.0，Ultra HD. The PET image reconstruction is performed using TureX+TOF, 2 iterations, and 21 subsets. Delayed scan is chest or pelvic PET/CT imaging performed again 60-120 minutes after the completion of trunk PET/CT acquisition. The remaining scanning parameters are the same as above.

### B. Experimental Setting

**Evaluation Metrics.** To evaluate the quality of image reconstruction, one perceptual metric, Fréchet inception distance (FID) and three typical metrics for evaluating image quality, including peak signal-to-noise (PSNR), structural similarity index (SSIM), and mean squared error (MSE) were employed to evaluate the performance of model. The PSNR and MSE are defined as follows:

$$\text{MSE} = \frac{1}{n}\sum_{n}^{i=1}(x_i - \hat{x}_i)^2 \quad (19)$$

$$\text{PSNR} = 20 \times \lg(\frac{MAX(x)}{\sqrt{MSE(x)}}) \quad (20)$$

where $x_i$ and $\hat{x}_i$ denote reconstruction image and ground truth image, respectively. PSNR and MSE focus on pixel-wise matching between ground truth image and reconstruction image. Therefore, they can evaluate the details of the reconstructed image well, but only partially evaluate the structure of the reconstructed image. SSIM is formulated as follows:

$$\text{SSIM}(x,y) = \frac{(2\mu_x\mu_{\hat{x}} + c_1)(2\sigma_{x\hat{x}} + c_2)}{(\mu_x^2 + \mu_{\hat{x}}^2 + c_1)(\sigma_x^2 + \sigma_{\hat{x}}^2 + c_2)} \quad (21)$$

it evaluates image quality in three aspects: brightness, contrast, and structure. However, both PSNR and SSIM can only evaluate the reconstructed images at the pixel level and statistical level, unable to truly reflect human observations of image quality. Dual-time PET imaging needs to play a role in clinical applications, so human perception of image quality is also an important evaluation metrics. Then we introduce FID, it is a Fréchet distance between the hidden feature map of reconstruct images and ground truth images which performing a pretrained inception-v3 model [49]. The define of Fréchet distance is described as follows:

$$\text{FID}^2 = \|\mu_h + \mu_{\hat{h}}\|^2 + Tr(\sigma_h + \sigma_{\hat{h}} - 2\sqrt{\sigma_h\sigma_{\hat{h}}}) \quad (22)$$

where $h$ and $\hat{h}$ denote the hidden feature map of reconstructed images and target images from pretrained inception-V3 model, $\mu$ denotes mean, $\sigma$ denotes covariance matrix and $Tr$ denotes calculate the trace of a matrix. Inception-V3 model is a powerful image classification network that can effectively extract image features, thus the hidden feature of inception-v3 can be seen as a model of image perception, which can to some extent approximate human perception of images.

**Implementation Details.** Our model was implemented by Pytorch framework and trained on an NVIDIA GeForce RTX 3060 Ti with 8G memory. The network training follows diffusion model manner for 300 timesteps and 1000 epochs using Adam optimizer which $\beta_1 = 0.5$, $\beta_2 = 0.999$, $\sigma_{lr} = 0.0001$ with a batch size of 8. The number of stages of model is 5, the base dimension of each stage is 32, the number of CNN block and transformer block in each stage are both two.

## V. RESULTS

To assess the capabilities of st-DTPM model, we conduct a comparative analysis against five prominent networks: DDPM, DiT, U-net, ResUnet, and Pix2Pix. We comprehensively analyzed the generation prediction capability of st-DTPM using two different datasets. The averaged quantitative results achieved are detailed in Table I in terms of four various evaluation metrics. Notably, st-DTPM achieves better results with higher PSNR and SSIM scores coupled with lower MSE and FID scores in both datasets. These findings underscore the ability of st-DTPM to generate delayed scan PET images of exceptional quality and precision, closely aligning with true delayed scan PET images. Meanwhile, in the comparison with DDPM generated results, we further confirm that the embedded transformer model and delay time variables can better capture the subtle changes and features in PET images, and the prediction results are more accurate. In contrast, DiT exhibits notably higher prediction errors, likely attributable to its diminished performance with limited datasets, necessitating extensive training for favorable outcomes. Besides, traditional generation models U-net, ResUnet and Pix2Pix are difficult to obtain the interaction between tissue structure and cancer tissue on the SUV values in different regions with only a single convolutional layer, so the generated images are blurry.

Fig. 5 presents compelling visual evidence through side-by-side comparisons of predicted delayed scan PET images alongside truly delayed images from three typical prostate cancer patients. It is worth to note that since the results of a single patient are presented here, we simply display the PSNR and SSIM of the images generated by each method.

Additionally, corresponding first scan PET images and truly delayed scan PET images are provided for thorough evaluation. These comparisons vividly demonstrate that our proposed method surpasses all competing approaches in terms of both image fidelity and preservation of structural details, highlighting its efficacy and advantages.

Despite encountering challenges related to tissue heterogeneity, particularly in lung predictions, our approach still achieves commendable average scores of 26.21 for PSNR, 0.8314 for SSIM, 23.94 for MSE, and 22.22 for FID. Qualitative insights from Fig. 6 further corroborate the better generative performance of our method compared to similar models, affirming its capability to produce credible and diverse PET images. In contrast, the generation results of other methods roughly restore the skeleton of real PET images while with blurry estimations in structural details.

TABLE I
PSNR, SSIM, MSE($*E^{-4}$), AND FID COMPARISON WITH DIFFERENT METHODS UNDER PROSTATE CANCER DATASETS AND LUNG CANCER DATASETS.

| Method | Prostate cancer dataset | | | | Lung cancer dataset | | | |
|---|---|---|---|---|---|---|---|---|
| | PSNR ↑ | SSIM ↑ | MSE ↓ | FID ↓ | PSNR ↑ | SSIM ↑ | MSE ↓ | FID ↓ |
| DDPM | 24.60 | 0.7534 | 34.65 | 39.42 | 21.56 | 0.5438 | 69.89 | 119.88 |
| DiT | 16.23 | 0.3018 | 238.50 | 347.01 | 15.51 | 0.1749 | 281.14 | 307.54 |
| U-net | 24.55 | 0.7337 | 35.04 | 79.09 | 24.10 | 0.7467 | 38.93 | 142.31 |
| ResUnet | 24.90 | 0.7737 | 32.37 | 248.53 | 23.92 | 0.7683 | 40.57 | 227.09 |
| Pix2Pix | 25.27 | 0.8250 | 29.75 | 45.44 | 22.67 | 0.7432 | 54.11 | 71.23 |
| **st-DTPM** | **28.80** | **0.8774** | **13.18** | **16.52** | **26.21** | **0.8314** | **23.94** | **22.21** |

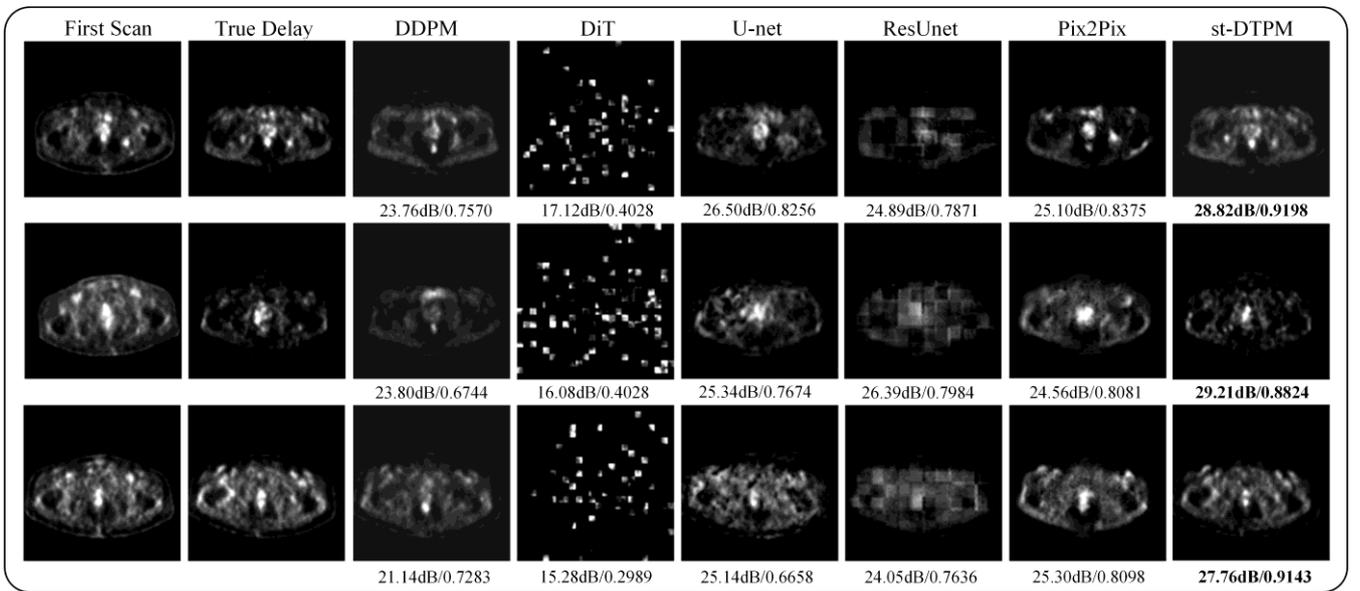

Fig. 5. Qualitative comparison of delayed scan PET images generated by various methods on prostate cancer datasets.

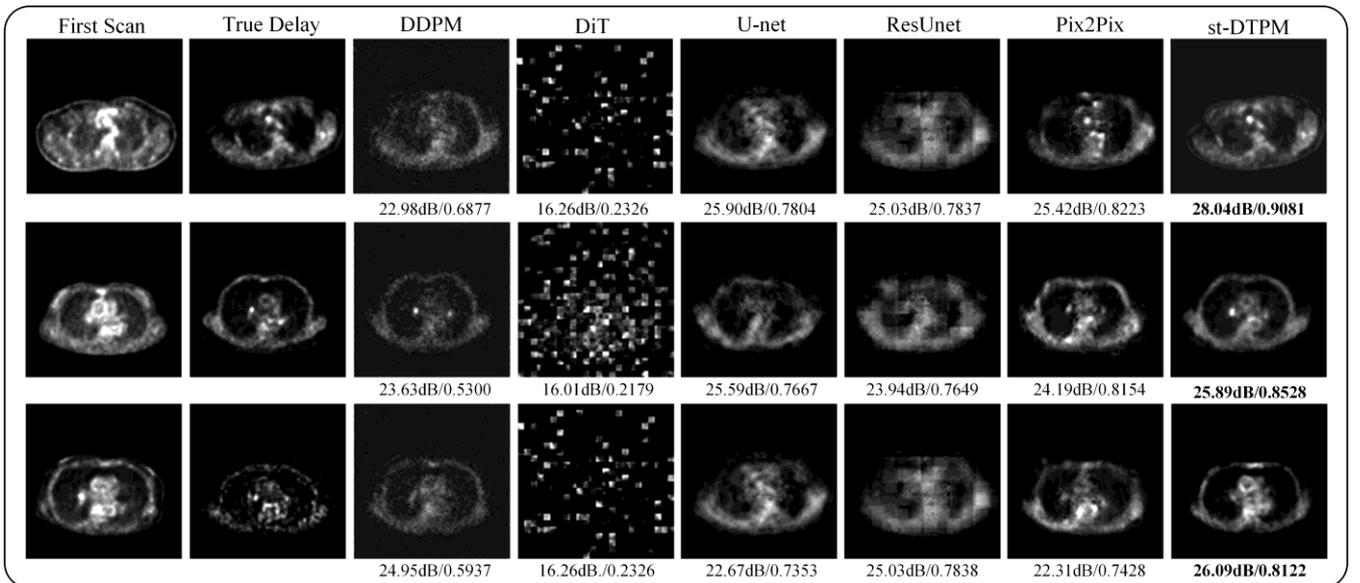

Fig. 6. Qualitative comparison of delayed scan PET images generated by various methods on lung cancer datasets.

In Section III. C, we explore two embeddings of universal time vectors. To investigate their interaction, we jointly embed them in our model and evaluate which combination yields optimal results. Table II displays four distinct embedding combinations. For instance, "EC" indicates that both modules embed time step vectors and delay time vectors separately; "LC" signifies concatenation followed by linear projection in the CNN module and transformer module. Similar patterns apply to subsequent combinations. From experimental results, while these four combinations are proposed based on our st-DTPM model, embedding both modules as "EC" consistently produces outstanding delayed scan PET images compared to alternative methods.

This phenomenon arises from potential information confusion or inconsistency during coupling due to covering distinct types of information by time step in diffusion models and delay time in delay imaging. Furthermore, convergence and divergence analyses are performed on these combinations to validate the superiority of "EC". Fig. 7 illustrates that utilizing "EC" results in more stable training processes and higher final PSNR compared to other methods. Based on the loss map of their training, it can be further inferred that the "EC" method is more effective. This is manifested in a faster rate of convergence and ultimately achieves the lower losses. As a result, all experiments related to st-DTPM in this study adopt the "EC" embedding mode.

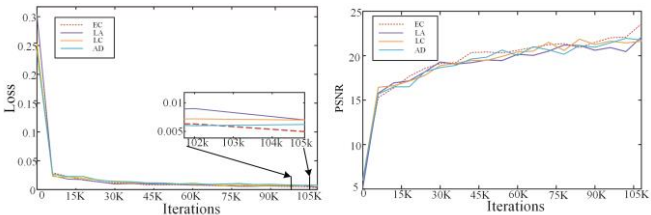

**Fig. 7.** Convergence of four embedding methods based on st-DTPM in terms of loss and PSNR from prostate cancer datasets.

TABLE II
PSNR, SSIM, MSE(*E$^{-4}$), AND FID COMPARISON WITH DIFFERENT EMBEDDING METHODS UNDER PROSTATE CANCER DATASETS.

| Method | PSNR ↑ | SSIM ↑ | MSE ↓ | FID ↓ |
|---|---|---|---|---|
| **EC** | **28.73** | **0.8762** | **13.40** | **15.74** |
| LA | 26.77 | 0.8468 | 21.06 | 21.81 |
| LC | 26.62 | 0.8394 | 21.77 | 21.78 |
| AD | 25.68 | 0.8185 | 27.05 | 27.89 |

In this section, we extensively discuss the ablation study of each key component proposed in our method. It is well known that the transformer model has demonstrated exceptional performance in natural language processing and other sequence modeling tasks. Its self-attention mechanism enables the model to effectively capture global information, thereby enhancing the accuracy of predictions. Similarly, incorporating temporal constraints such as delay time embeddings further improves predictive precision.

*Contribution of the Transformer.* We compare the performance of the baseline DDPM with the DDPM + Transformer (DDPM + Trans) model to validate the effectiveness of the transformer architecture. Quantitative comparisons of PSNR and SSIM are detailed in Table III, revealing significant improvements across all metrics for DDPM + Trans manner compared to the original DDPM on the same dataset with PSNR increasing by 2.64dB and SSIM by 0.0504. Furthermore, embedding the transformer model can make the generated delayed scan PET images closer to the real images with the lower MSE and FID. These experimental results clearly demonstrate that integrating the transformer architecture significantly enhances the quality of generated images, highlighting its importance in our model.

*Contribution of the Delay Time Embedding.* To investigate the efficacy of delay time embedding, we also compare the performance of the baseline DDPM with the DDPM + delay time embedding model (DDPM + DT). Without delay time embedding, PSNR and SSIM values are 25.99 and 0.8267 respectively, whereas incorporating delay time embedding raised average PSNR to 27.41 and SSIM to 0.8500 under the same conditions. At the same time, it also achieves lower MSE and FID. These quantitative results visibly indicate that adding delay time embedding yields higher synthesis quality in generated images, underscoring the importance of this technique.

Furthermore, quantitative comparisons among the latter three demonstrate that DDPM + Transformer + delay time embedding (DDPM + Trans + DT) way achieves outstanding performance across most metrics. This further emphasize the effectiveness of both transformer architecture and delay time embedding in our model.

TABLE III
PSNR, SSIM, MSE(*E$^{-4}$), AND FID COMPARISON WITH DIFFERENT HYBRID NETWORKS UNDER PROSTATE CANCER DATASETS.

| Method | PSNR ↑ | SSIM ↑ | MSE ↓ | FID ↓ |
|---|---|---|---|---|
| DDPM | 25.99 | 0.8262 | 25.20 | 31.62 |
| DDPM + Trans | 28.63 | 0.8766 | 13.72 | **14.99** |
| DDPM + DT | 27.41 | 0.8530 | 18.15 | 27.04 |
| DDPM + Trans+ DT | **28.81** | **0.8776** | **13.15** | 15.51 |

## VI. DISCUSSION

This work proposed a new dual-time PET delay imaging method named st-DTPM. The core idea of this method was to generate the PET images of the first scan through a network model, so as to realize the prediction of the delayed scan PET images. To match such a model, we leveraged the powerful generation capabilities of DDPM to generate delayed scan PET images. First, we used early scanned PET images and noisy delayed scanned images to train a deep noise reduction model to learn the prior information between the images. This process involved capturing complex spatial and temporal relationships in the data to improve the accuracy of predictions. Secondly, we embedded transformer structure to U-net denoising network inside DDPM to form a hybrid noise reduction model. At the same time, we converted the delay time variable and time step into a general vector and embed it into the network to further improve the prediction accuracy of the model.

Our proposed st-DTPM model demonstrated better predictive accuracy compared to other generative models. To comprehensively assess its performance, we conducted comparative analyses across prostate cancer and lung cancer datasets, evaluating predictions generated by U-net, ResUnet, Pix2Pix, DiT, and st-DTPM. Specifically, when only using the U-net network, it failed to capture the mutual influence between tissue structures and cancerous tissues on the SUV values across different regions. Consequently, both U-net and ResUnet models produce less accurate predictions than st-DTPM. At the same time, employing transformer alone in the model construction lacked the necessary inductive bias to effectively reconstruct the intrinsic organizational features of PET images, particularly when data was limited. This limitation resulted in larger prediction errors for PET delay using DiT. By using the hybrid conditional diffu-sion model method of st-DTPM, we achieved significantly improved model prediction performance. This method enhanced accuracy by capturing intricate spatial-temporal relationships and effectively incorporated temporal information, thereby overcoming the limitations of other models.

The paired PET images from dual-time were not registered. However, st-DTPM still has accurate predictive performance. This demonstrates its robustness, as it can partially correct image position offsets while learning image mapping relationships.

While the st-DTPM model has demonstrated excellent results in these comparative experiments, there are still notable limitations that require addressing. A primary concern is whether the inherent low resolution of PET images may introduce interference in delay prediction. This issue stems from the inherent limitations of PET imaging technology. Moving forward, we will focus on developing more accurate PET delay prediction methods to mitigate these challenges in future studies.

## VII. CONCLUSION

In this work, we introduced a hybrid diffusion transformer model designed for predicting delayed scan PET images. Inspired by the DDPM architecture, our approach utilized early delayed scan PET images and delayed scan PET images with added noise as inputs, employing the hybrid diffusion transformer model to denoise and generate images from standard Gaussian noise. Experimental results demonstrated that the st-DTPM method achieved better performance metrics including higher PSNR and SSIM, lower MSE and FID compared to alternative models and excelled in generating high-quality delayed scan PET images.

## ACKNOWLEDGMENT

The authors sincerely thank the anonymous referees for their valuable comments on this work. All authors declare that they have no known conflicts of interest in terms of competing financial interests or personal relationships.